\documentclass[10pt]{article}
\usepackage{subfigure,epsfig,multicol}
\begin{document}

{\bf \large The Hydride Anion in an Extended Transition Metal
Oxide Array: LaSrCoO$_3$H$_{0.7}$}

M.A. Hayward\footnote{Department of Chemistry, University of Liverpool,
Liverpool,  L69  7ZD, U.K.}, E.J. Cussen$^1$, J.B. Claridge$^1$, M.
Bieringer$^1$, 
M.J.  Rosseinsky$^1$*, C.J. Kiely$^{1,}\footnote{Materials Science
Division, Department  of  Engineering, University  of Liverpool, L69
3BX, U.K.}$, S.J. Blundell\footnote{Clarendon Laboratory,  Department
of  Physics, University of  Oxford, Oxford, OX1 3PU, U.K.},\newline
I.M. Marshall$^3$ and F.L. Pratt\footnote{RIKEN-RAL  Muon  Facility,
Rutherford  Appleton  Laboratory,  Chilton, Didcot, Oxon, OX11 0QX, U.K.}

{\bf We present the synthesis and structural characterisation of a
transition metal oxide hydride, LaSrCoO$_3$H$_{0.7}$, which adopts an
unprecedented structure in which oxide chains are bridged by hydride
anions to form a two- dimensional extended network. The metal centers
are strongly coupled by their bonding with both oxide and hydride
ligands to produce magnetic ordering up to at least 350\,K. The
synthetic route is sufficiently general to allow the prediction of a
new class of transition metal-containing electronic and magnetic
materials.}

\begin{multicols}{3}
The covalent interaction between the O$^{2-}$ anion and the d-orbitals
of the transition metal cation is at the heart of the remarkable
electronic properties of the transition metal oxides(1, 2): even in
mixed-anion oxyhalides(3), it is the metal-oxide interactions which
dominate the physical properties. Developing synthetic routes to
materials in which other anions partially replace oxide could open up
the possibility of preparing entirely novel families of electronically
active transition metal compounds.  The hydride anion, H$^-$, with a
1s$^2$ electronic configuration, is known to engage in strong covalent
bonding with transition metal centres in discrete molecular species(4)
and would be an excellent candidate for the transmission of exchange
interactions or electron delocalization between transition metal
cations in an oxide hydride, if the formidable synthetic difficulties
barring the path to such a phase could be overcome.  The problem is
that H$^-$, unlike O$^{2-}$ or halide anions such as F$^-$ and Cl$^-$,
is a powerful reducing agent and would be expected to transform the
transition metal component of a typical high-temperature ternary
transition metal oxide synthesis into the metal, defeating most
possible synthetic strategies. Here we demonstrate a low-temperature
topotactic route to insert H$^-$ anions directly into an extended
transition metal oxide array, and show that H$^-$ transmits exchange
interactions between the transition metal cations at least as
effectively as O$^{2-}$, opening up a new mechanism for designing
co-operative effects in solids.

We have recently shown that NaH is an effective low-temperature
reducing agent for ternary transition metal oxides.  At temperatures
below 190$^\circ$C NaH affords the Ni(I)(5) and Co(I)(6) oxidation
states, but at higher temperatures completely reduces the metal
because of the presence of hydrogen gas in thermal equilibrium with
the hydride salt.  In order to study the solid-state reactivity of
H$^-$ with ternary transition metal oxides at higher temperatures, we
used the more thermally stable CaH$_2$ (with a decomposition temperature
of 885$^\circ$C compared with 210$^\circ$C for NaH). CaH$_2$ was
reacted with the Co(III) oxide LaSrCoO$_4$, which adopts the layered
K$_2$NiF$_4$ structure with square planar CoO$_2$ sheets alternating
with (La/Sr) O rock-salt layers and octahedral co-ordination around
Co(III).  Reaction for two periods of 4 days at 450$^\circ$C in a
sealed Pyrex tube with intermediate grinding afforded a mixture of CaO
and an orthorhombic phase {\bf 1} (7).  The orthorhombic phase is
structurally related to the starting material and its lattice
parameters suggest a one-dimensional Sr$_2$CuO$_3$ structure that
consists of chains of corner sharing MO$_4$ squares (7). The
transformation to such a structure suggests that {\bf 1} is the Co(I) phase
LaSrCoO$_3$, formed by the reductive topotactic extraction of O$^{2-}$
to afford CaO.  CaO was removed from {\bf 1} by washing with 0.1M NH$_4$Cl
in degassed methanol under nitrogen and then filtering and drying
under vacuum.  The structure and composition of {\bf 1} were investigated by
powder synchrotron X-ray diffraction (XRD) and neutron diffraction,
and selected-area electron diffraction (SAED).

The 290\,K synchrotron XRD pattern was readily indexed with the
orthorhombic Sr$_2$CuO$_3$-type unit cell in the {\it Immm} space group, an
assignment confirmed by SAED (7).  Structural refinement showed that
the cations occupy the metal positions expected for the Sr$_2$CuO$_3$
structure.  The ambient temperature neutron powder diffraction data,
however, could not be indexed on this basis; a doubling of the basal
{\it ab} plane area was required to account for additional diffraction
reflections observed at large {\it d}-spacings. The absence of these
reflections in the XRD and SAED data, which would be sensitive to
superstructure formation due to chemical or crystallographic ordering,
suggests they are of magnetic origin.(5)

Magnetic long-range order in a strongly one-dimensional (1D) structure
such as that adopted by Sr$_2$CuO$_3$ is totally unexpected at 300\,K.
We therefore recorded muon spin rotation ($\mu$SR) data(8) to investigate
the behavior of the magnetic moments carried by the cobalt
cations(Fig.\ 1A).  Clear oscillations are apparent over the entire
temperature range which signify a quasi-static magnetic field at the
muon site.  This result demonstrates unambiguously that {\bf 1} is uniformly
magnetically ordered throughout its bulk at temperatures up to at
least 310\,K.  The amplitude of the oscillations corresponds to a
signal from the whole sample so these results exclude the possibility
that only a small region of the sample is ordered.  The frequency of
the oscillations approaches 71\,MHz as T$\to$0 (corresponding to an
internal field of 0.53\,T) and decreases as the sample is warmed to
310\,K (Fig.\ 1C).  At the highest temperatures the relaxation rate of
the oscillations began to increase and probably reflects the approach
of the phase transition. (Fig.\ 1B). These results do not allow us to
determine precisely the N\'eel temperature, but show that it is likely
to be above 350\,K. The muon site is likely to be similar to that
found in Sr$_2$CuO$_3$ and Ca$_2$CuO$_3$, in which the muon is
believed to form an O--$\mu^+$ bond with a bond length of $\sim$1~\AA.(9, 10)
However the measured muon precession frequency for these materials is
much lower, corresponding to 2.32 and 3.5~mT for Sr$_2$CuO$_3$ and
Ca$_2$CuO$_3$ respectively (10). Such low internal fields are
associated with magnetic moments less than 0.1~$\mu_{\rm B}$.  The
precession frequency in the present case is $> 200$ times greater than
in Sr$_2$CuO$_3$, which is consistent with the moment refined from
powder neutron diffraction on {\bf 1}. The N\'eel temperatures of
Sr$_2$CuO$_3$ and Ca$_2$CuO$_3$ are $\sim 5$\,K and $\sim 11$\,K
respectively.  The much larger value of $T_{\rm N}$ 
observed for {\bf 1} points to a
large interchain coupling, which is surprising because the lattice
constants in the interchain direction are larger than in Sr$_2$CuO$_3$
and Ca$_2$CuO$_3$.

After the verification of long-range magnetic order at room
temperature by $\mu$SR, a simple antiferromagnetic (AF) ordering model,
with antiparallel spin alignment of all neighboring Co spins, was
incorporated in refinement of the neutron data.  Although this
addition gave a satisfactory fit to the magnetic reflections, the fit
to the nuclear reflections was unsatisfactory ($\chi^2 = 5.15$, 
weighted profile residual ($R_{\rm wp}$) = 5.55\%) (7).  
A difference Fourier map was computed and revealed a strong
peak of negative scattering density at (0, 1/2, 0) midway between the
Co atoms along the b-axis at the vacant oxide anion position in the
CoO$_{2-x}$ sheets (7). Hydrogen is one of the few elements to have a
negative neutron scattering length,(11) and therefore hydrogen was
inserted into the model at this position. The refinement immediately
converged at $\chi^2 = 1.96$, $R_{\rm wp}$ = 3.43\% as shown in Fig.\ 2,
demonstrating that {\bf 1} is the first extended transition metal oxide
hydride.

The structural analysis was completed by a three histogram refinement
of 2.4~\AA\ and 1.59~\AA\ neutron histograms together with the
synchrotron data (Fig.\ 2). The quality of the fits demonstrates the
correctness of the structural model, which gives a refined composition
of LaSrCoO$_3$H$_{0.70(2)}$.  The presence of a small amount of
La$_2$O$_3$ is not inconsistent with this composition (7). The ordered
magnetic moment carried by the cobalt cations increases from
1.77(5)$\mu_{\rm B}$
at 290\,K to 1.95(4)$\mu_{\rm B}$ at 2\,K.  The +1.7 oxidation state of Co
deduced from the refined composition with a charge distribution
assigning the $-1$ oxidation state to H is consistent with the
position of the Co K-edge in LaSrCoO$_3$H$_{0.70(2)}$,(7) and the
presence of hydride was confirmed chemically by quantitative mass
spectrometric monitoring of the H$_2$O evolved simultaneously with
oxidation of {\bf 1} under flowing O$_2$ at 272(5)$^\circ$C, which indicates
0.4 H$^-$ per formula unit; while necessarily less accurate than the
structure refinement, this chemically confirms the presence of
hydrogen.  Chemical analysis (5) reveals 0.27\% H by mass compared
with 0.21\% expected for LaSrCoO$_3$H$_{0.70(2)}$.

The structure (Fig.\ 3) consists of chains of CoO$_4$ squares sharing
corners along a to form chains which are linked into a 2D array in the
$ab$ plane by H$^-$ bridges along $b$.  
The CoO$_2$ sheets in the ab plane
of the starting material have been replaced with CoOH$_{0.7}$ sheets
in the oxide hydride product, consistent with a reduction-insertion
mechanism in which oxide vacancies are created in the xy plane
followed by their filling by the H$^-$ anions.  The Co(II) cations
have a mean coordination number of 5.40(4), with the H$^-$ anions
occupying the axial positions between the square plane of oxide
anions.  The Co--H distance of 1.80174(2)~\AA\ is shorter than either of
the Co--O distances and this, coupled with the strong covalency
expected for the interaction between Co$^{2+}$ and H$^-$, produces
strong AF coupling between the Co(II) cations within the 2D
sheets. The effect of the hydride anions on bridging the
Sr$_2$CuO$_3$-like 1D chains is qualitatively demonstrated by
comparing $T_{\rm N}$ of above 300\,K here with 11\,K in Sr$_2$CuO$_3$ itself.
A qualitative comparison between H$^-$ and O$^{2-}$oxide bridges can be
made by noting that LaSrCoO$_{3.5}$ (6) has a N\'eel temperature of
110\,K, with a similar concentration of bridging anions in the $ab$
plane.

The high AF ordering temperature of LaSrCoO$_3$H$_{0.70}$ demonstrates
that the H$^-$ anion can strongly couple transition metal cations
electronically.  To confirm this supposition, we have performed spin
wave calculations to crudely model the effect of coupling the CoO$_3$
oxide chains by H$^-$ anions.  We find that increasing the effective
exchange between chains in one direction to close to the intrachain
exchange $J$ is sufficient to raise the N\'eel temperature to the order
of $J/k_{\rm B}$ (where $J$ 
is the intrachain exchange), even if the coupling in
the orthogonal interchain direction is quite weak; the size of the
moment will then be near the full value (this scenario is appropriate
for LaSrCoO$_3$H$_{0.70}$, in which we observe a large moment and
$T_{\rm N}\sim 350$~K which is of the order of $J/k_{\rm B}$).  If, however 
the interchain
exchange is weak in both directions, the N\'eel temperature is largely
controlled by that weak interchain exchange and the size of the moment
is greatly reduced because of quantum fluctuations (this scenario is
appropriate for Sr$_2$CuO$_3$, in which the moment is reduced to
0.06\,$\mu_{\rm B}$ and $T_{\rm N}= 11$~K (8)). This stark difference in
magnetic properties highlights the crucial role played by the bridging
H$^-$ ions, and confirms that they are capable of coupling transition
metal centres equally as effectively as O$^{2-}$ anions.

The H$^-$ anions are exclusively located in the transition metal
containing layers in the structure, in contrast to oxyhalides such as
Sr$_2$CoO$_3$Cl(3) in which the halide anions occupy the apical
positions within the electronically inactive rock salt layers. The
demonstration that transition metal oxide hydrides can be isolated now
suggests such species should be considered when discussing chemical
and catalytic processes involving transition metal oxides, such as the
spillover process(12) where only H$^{+}$ and OH$^-$ have been
invoked.(13) The site-specific insertion of the H$^-$ anions into the
oxide sheets may reflect the enhanced bridging interaction with the
transition metal cations in this site, or be due to a topotactic
transformation mechanism in which anion vacancies present in an
LaSrCoO$_{3.5}$ intermediate are filled with hydride anions.  The
combination of cation reduction with anion insertion is unusual for a
topotactic solid-state transformation, but may prove to be a general
route to transition metal oxide hydrides, opening up previously
uncharted areas in electronic and magnetic materials
synthesis. Quantitative estimates of the strength of the exchange
interaction demonstrate that the H$^-$ bridge couples metal centres as
effectively as O$^{2-}$, although the different frontier orbital
symmetry ($\sigma$ only in H$^-$, $\sigma$ + $\pi$ in O$^{2-}$)
promises interesting property differences to be revealed in future
detailed studies of this new class of extended solid.(14)

{\bf References and notes}

1.  J. B. Goodenough, J. S. Zhou, {\em Chem. Mater.} {\bf 10}, 2980
(1998).\newline
2.  Y. D. Chuang, A. D. Gromko, D. S. Dessau, T. Kimura, Y. Tokura,
{\em Science} {\bf 292}, 1509 (2001).\newline
3.  S. M. Loureiro, C. Felser, Q. Huang, R. J. Cava, {\em
Chem. Mater.} {\bf 12}, 3181 (2000).\newline
4.  Z. Y. Lin, M. B. Hall, {\em Coord. Chem. Rev.} {\bf 135}, 845
(1994).\newline
5.  M. A. Hayward, M. A. Green, M. J. Rosseinsky, J. Sloan, {\em
J. Am. Chem. Soc.} {\bf 121}, 8843 (1999).\newline
6.  M. A. Hayward, M. J. Rosseinsky, {\em Chem. Mater.} {\bf 12}, 2182
(2000).\newline
7.    Full synthetic and structural characterization details are provided
as supplemental data at Science online.\newline
8.  S. J. Blundell, {\em Contemporary Physics} {\bf 40}, 175
      (1999).\newline
9.  A. Keren, et al., {\em Physical Review B} {\bf 48}, 12926
(1993).\newline
10.   K. M. Kojima, et al., {\em Physical Review Letters} {\bf 78}, 1787
      (1997).\newline
11.   G. E. Bacon, Neutron Diffraction (Clarendon Press, Oxford, ed. 3rd,
1975).\newline
12.  W. C. Conner, J. L. Falconer, {\em Chem. Rev.} {\bf 95}, 759
(1995).\newline
13.  S. J. Teichner, {\em Appl. Catal.} {\bf 62}, 1 (1990).\newline
14.  We thank the EPSRC for support under GR/N21819 and for access to
ILL and ESRF.  We thank the donors of the Petroleum Research
Foundation, administered by the American Chemical Society, for
support.  We thank T.\ Hansen (ILL) and A.\ Fitch (ESRF) for their
expert assistance with the collection of neutron and synchrotron X-ray
diffraction data, the staff of the PSI muon facility for their
assistance with the muon experiments and L.\ Murphy (SRS) for
assistance with the collection of X-ray absorption data.  We are
grateful to R.\ Coldea for useful discussions.
\end{multicols}

\begin{figure}
\centerline{
\psfig{figure=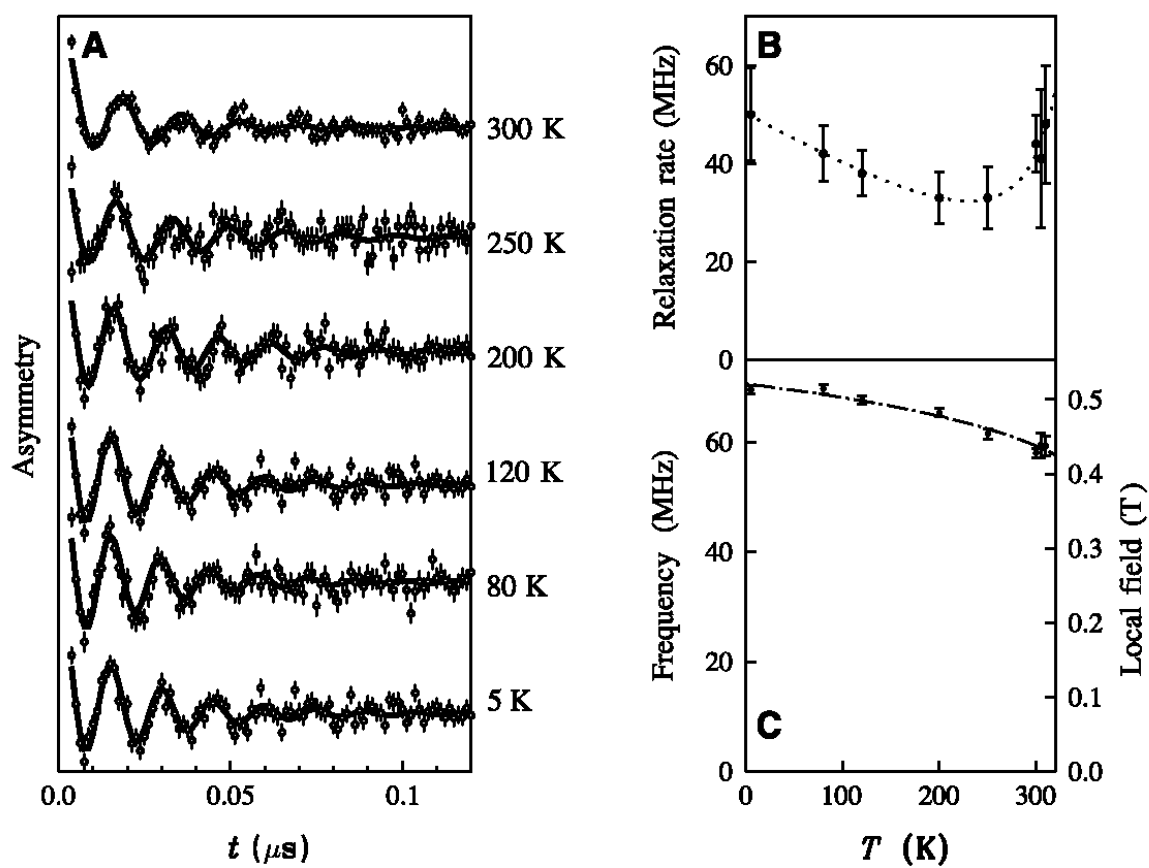,width=0.9\textwidth}}
\caption{(A) Muon spin rotation data from {\bf 1}:  The  oscillations  in
          the  asymmetry  demonstrate  long-range  magnetic  order  at   all
          temperatures measured (up to 310~K) temperature dependence  of  (B)
          the relaxation rate of the oscillations (dotted line is a guide to
          the  eye)  and  (C)  the  muon-spin  rotation  frequency  and  the
          corresponding magnetic field at the muon site.  In  (C)  the  line
          represents  a  fit  to  a  phenomenological  expression  for   the
          temperature dependence  of  the  order  parameter  in  a  magnetic
          material.  Good fits could be obtained by  fixing  the  transition
          temperature at values in the range 350  to  450~K,  demonstrating
          that the magnetic ordering temperature of {\bf 1}  probably  lies  above
          350~K.}
\end{figure}

\begin{figure}
\centerline{
\psfig{figure=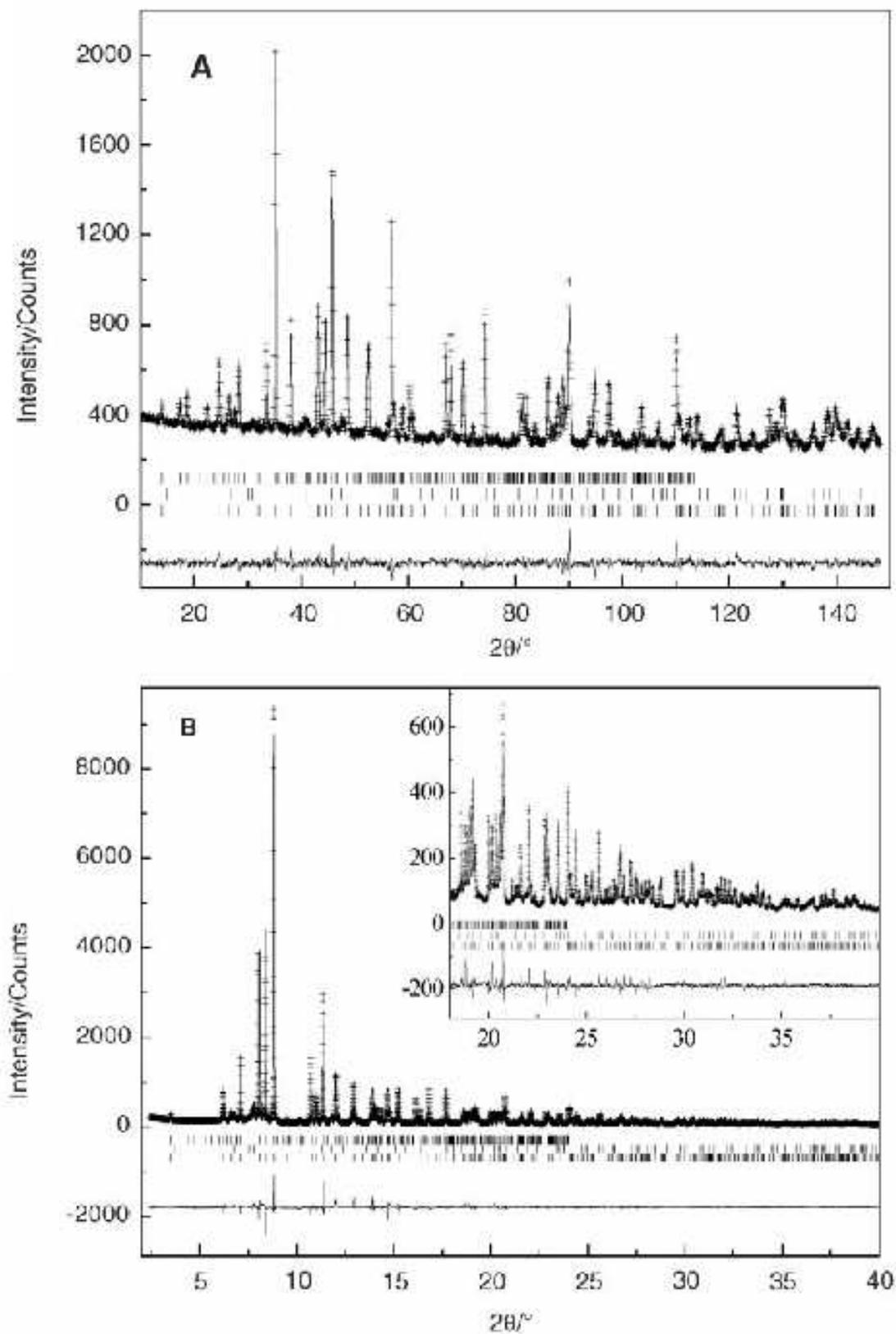,width=0.9\textwidth}}
\caption{Structural  characterisation  of {\bf 1}, LaSrCoO$_3$H$_{0.70}$   by
          simultaneous Rietveld refinement of (A)  neutron  diffraction  and
          (B) synchrotron XRD data collected as described in (5), where full
          refinement details are also available. {\bf 1} adopts space group  {\it Immm},
          a = 3.87093(4)\AA, b  =   3.60341(3)  \AA,  c  =
          13.01507(10) \AA,  V  =
          181.541(3) \AA$^3$, La (50 \%  occupancy)/Sr  (50  \% 
          occupancy)  on  4i 
          0,0,0.35703(3), Co on 2a 0,0,0, O(1) on 4i 0,0,0.1673(2),  O(2)  on
          2b 1/2, 0, 0, H on 2d 0, 1/2,0 70(2) \% occupancy.}
\end{figure}

\begin{figure}
\centerline{
\psfig{figure=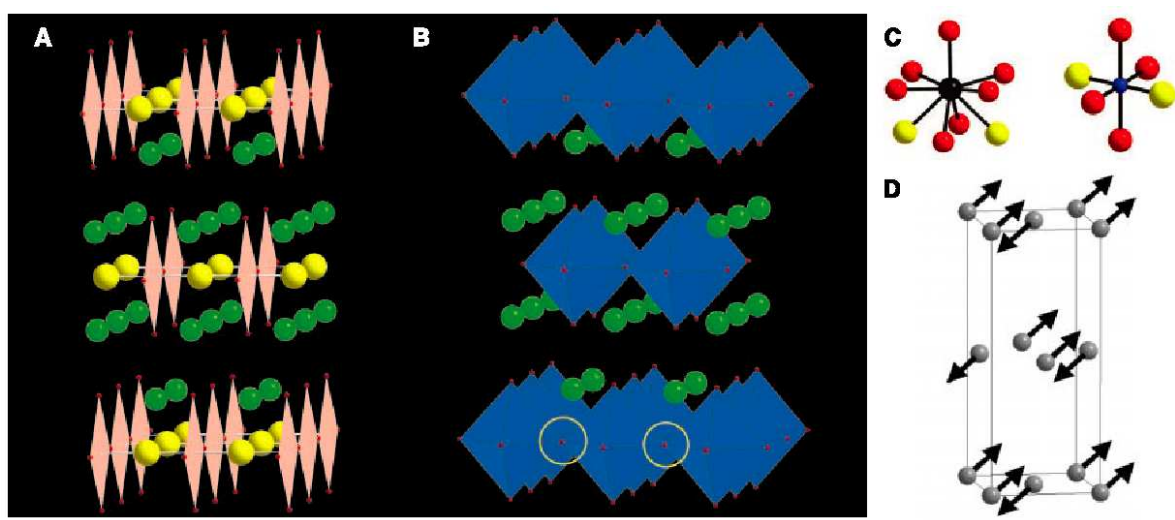,width=0.9\textwidth}}
\caption{(A) The crystal  structure  of  {\bf 1}, LaSrCoO$_3$H$_{0.70}$  and  its
          topotactic relation with (B) the LaSrCoO$_4$ starting
          material. {\bf 1} can be generated from LaSrCoO$_4$ by
          substitution of half of the equatorial O$^{2-}$ anions (circled
          in yellow) by H$^-$. Hydride ions and Sr/La cations are
          represented as yellow spheres and green spheres respectively
          and the oxide ion arrangements around the Co cations are
          illustrated by squares and octahedra in (A) and (B)
          respectively.  Co--O(1) 2.1779(22) \AA\ $\times$ 2, 
          Co--O(2) 1.93549(2) \AA\ $\times$ 2, Co--H 1.80174(2) \AA\
          $\times$ 2, Sr/La--O(1) 2.469(2) \AA, 2.6633(3) \AA\
          $\times$ 4, Sr/La-O(2) 2.5902(3) \AA $\times$ 2, Sr/La--H
          2.6849(3) \AA\ $\times$ 2 The bond angles within the
          CoO$_4$H$_{1.4}$ unit are all constrained to 90$^\circ$ or
          180$^\circ$ by the symmetry of {\bf 1}.  (C) The coordination
          environments of the La/Sr (black sphere) and Co (blue
          sphere) sites in {\bf 1}.  Oxide and hydride anions are
          represented as red and yellow spheres, respectively.  (D)
          The magnetic structure of {\bf 1} at room temperature.  For
          increased clarity, only the Co cations are shown.  The
          arrows represent the direction of the ordered magnetic
          moments.}
\end{figure}

\end{document}